\documentclass[12pt]{article}

\def\be{\begin{equation}}
\def\ee{\end{equation}}
\def\bea{\begin{eqnarray}}
\def\eea{\end{eqnarray}}

\def\be{\begin{equation}}
\def\ee{\end{equation}}
\def\bea{\begin{eqnarray}}
\def\eea{\end{eqnarray}}
\def\half{\frac{1}{2}}
\def\nn{\nonumber}

\def\case#1/#2{\textstyle\frac{#1}{#2}}
\def\k0{\kappa_{0}}

\begin{document}
\begin{titlepage}

\vspace{.7in}

\begin{center}
\Large
{\bf Some Aspects of  Pre Big Bang Cosmology}\\
\vspace{.7in}
\normalsize
\large{ 
Alexander Feinstein
}\\
\normalsize
\vspace{.4in}

{\em Dpto. de F\'{\i}sica Te\'orica, Universidad del Pa\'{\i}s Vasco, \\
Apdo. 644, E-48080, Bilbao, Spain}\\
\vspace{.2in}

\vspace{.3in}
\baselineskip=24pt
\begin{abstract} 
This is a summary of a course given at the Fourth Mexican School 
on Gravitation and Mathematical Physics on some aspects of PBB cosmology. 
After  introductory remarks the lectures concentrate on some amusing
consequences derived from the symmetries of the string theory with respect 
to such classical  concepts as isotropy and homogeneity. The  extra dimensions and the symmetries
of the M theory  are further applied to show that the classical singularities might
be just physically irrelevant. In the final lecture a  model universe is ``produced" from ``almost nothing"
and it is argued that initial plane waves are thermodinamically natural state for the universe to emerge from.
\end{abstract}

\vspace{3in}
\it{Lectures given at the Fourth Mexican School on Gravitation and Mathematical Physics, 
Huatulco, Mexico 2000}

\vspace{.2in}
\end{center}

\end{titlepage}

\section{Itroduction}

First I would like to thank the Organizers for giving us the opportunity to give lecture Course
in such a beautiful surrounding. 

As far as my course is concerned, I will be  concentrating on our recent works in String Cosmology
done mostly in collaboration with Kerstin Kunze and Miguel V\'azquez-Mozo.

I will first briefly review some general aspects of String Cosmology, and in the following days we will
touch  the Initial Conditions, and the Singularity Problem.
Most of the References may be  found in a nicely
organized M. Gasperini WEB page: http://www.to.infn.it/$\sim$gasperin.

The central problems of modern theoretical cosmology are: the initial conditions, the singularity problem
and the dimensionality of the universe.  The first to begin modern discussions of the initial conditions
problem was Ch. Misner \cite{misner} in his {\it Chaotic Cosmology Program.} The idea was to allow the universe
to start from an arbitrary complex initial state and to identify the mechanism to smoothen out 
the irregularities, to finally
explain as to why the universe is homogeneous and isotropic to such
a high extent. Unfortunately, even if one focuses on homogeneous but anisotropic cosmological
models, the so-called Bianchi models, only  those  which include the FRW models as particular cases
would isotropise, and these we say are of measure 0. 

{\it Inflation} \cite{guth}, somehow brought Misner's idea back to life, in the sense that perturbations
around FRW backgrounds were to disappear if the universe to undergo a period of accelerated expansion.
Nevertheless, generically non-linear inhomogeneities such as strong shock primordial waves etc
remain difficult to deal with.

A solution to the i.c. problem may come from  rather  different a direction, such as thermodynamics, for
example, if some idea as to how to assign entropy to gravitational field were available. In the 70's Penrose
\cite{penrose} put forward a 
speculation that one should assign a 0 gravitational entropy to geometries with 0 Weyl tensor, so that
the universe must have been  FRW to start with. Yet, this notion of
gravitational entropy is apparently unrelated 
to the notion of entropy even in the cases we know of, such as Black Holes, or entropy for 
high frequency gravity waves etc. Moreover, the idea of assigning 0 entropy to FRW geometries is influenced
strongly by a classical view on the initial state of the universe. Taking for example ``stringy" approach
to the early universe one would have rather thought that the background geometry should represent
a target space of some exact Conformal Field Theory, and FRW doesn't seem to be an exact CFT. Also there is
 rather a different physical view on matters such as isotropy and homogeneity in the context of the
string theory. I will be arguing later that certain kinds of inhomogeneities and anisotropies do not
really matter when symmetries of the low energy string theory are considered--thus it could be
that our notions of isometries must be changed.

The {\it PBB} \cite{gabr} cosmology is rather different. First, it pushes the i.c. into the ``cold radiative" past where
physics is believed to be known to us. The hot regime, near the singularity, now becomes the intermediate phase and it is
believed that the symmetries of the non-perturbative string theory may help to deal with
the singularity. To summarize the PBB scenario in one paragraph:

The universe starts as a perturbation in an otherwise flat background \cite{inhom},
\cite{buon}, \cite{fkvm1}, \cite{gas} this perturbation collapses from the point
of view of an Einstein observer. It expands and inflates in the so-called String frame. The inflation
drives the universe towards the ``intermediate" singularity meanwhile getting rid of all the
inhomogeneities and anisotropies and the string symmetries act to squeeze the universe unharmed
through the hot regime. Finally, emerges the observed universe.

The inflation in the PBB universe is driven by the kinetic energy of massless dilaton field which is the fundamental ingredient
of any string theory, and there are substantial differences between the standard inflation and the
PBB phase. In  standard inflation  the i.c. are prescribed in the regime one knows very little about
physics, the initial curvature scale may be very large and  remains constant
or decreases upon expansion, on the other hand, the initial coupling is arbitrarily weak in PBB,
the curvature grows while the initial curvature is practically 0 \cite{gas}.

Now, people often talk about String Cosmology when talking about PBB. What does one really mean when String Cosmology
is quoted? Basically we lack a complete non-perturbative string theory valid at Plank time. So really
the game is restricted to the energies $\sim 10^{19}$ GeV $\sim GUT$ $\sim 10^{16}$ GeV. Therefore,
one may write the low energy effective action for the theory which reduces to GR + a bunch of extra massless
fields depending on the type of the string theory. The questions, however, the cosmologists ask
remain the same: why the universe is homogeneous, what happens with initial singularity,
can we solve the horizon and flatness problem, and... does the universe we speculate about
 corresponds  to the one we observe?

On general grounds, the massless bosonic
sector of superstrings includes the gravitational field $G_{\mu\nu}$, 
the dilaton $\phi$, with vacuum expectation value determining the string 
coupling constant, and the antisymmetric rank-two tensor $B_{\mu\nu}^{(1)}$. 
There are more massless fields depending on the particular superstring model. The lowest 
order effective
action for the massless fields can be written as
$$
S_{\rm eff}={1\over (\alpha^{'})^{D-2\over 2}}
\int d^{D}x\sqrt{G}e^{-2\phi}\left[R+4(\partial\phi)^{2}-
{1\over 12}(H^{(1)})^{2}\right]+S_{\rm md},
$$
where $H^{(1)}=dB^{(1)}$ is the field strength 
associated with the NS-NS two-form
and $S_{\rm md}$ is a model-dependent part which includes other massless 
degrees of freedom.

	When $D<10$ some of these massless fields correspond
to gauge and moduli fields associated with the specific compactification 
chosen, and the dilaton $\phi$ appearing in the  above equation is related to the 
ten-dimensional
dilaton $\phi_{10}$ by $2\phi=2\phi_{10}-\log V_{10-D}$, where $V_{10-D}$ 
is the volume of the
internal manifold measured in units of $\sqrt{\alpha^{'}}$.

 Let us restrict the attention to generic degrees of freedom, leaving aside the 
internal components of the ten-dimensional fields.
In the heterotic string case, $S_{\rm md}$ 
contains the Yang-Mills action for the 
background gauge fields $A_\mu^{a}$.
In the case of the model-dependent part of the type-IIB superstring 
things are  more involved; among the
massless degrees of freedom in the R-R sector we find, 
along with a pseudo-scalar
$\chi$ (the axion) and a rank-two antisymmetric tensor $B_{\mu\nu}^{(2)}$, 
a rank four self-dual form $A_{\mu\nu\sigma\lambda}^{\rm sd}$. The 
presence of this self-dual form spoils the covariance of the effective action  
for the massless fields,
since there is no way of imposing  the self-duality condition in a generally 
covariant
way. But, if we set $A^{\rm sd}$ to zero, we can write a covariant action for 
the remaining fields with
$$
S^{\rm IIB}_{\rm md}=-{1\over (\alpha^{'})^{D-2\over 2}}
\int d^{D}x\sqrt{G}\left[{1\over 2}(\partial\chi)^{2}+
{1\over 12}(H^{(1)}\chi+H^{(2)})^{2}\right]
$$
Notice that the R-R fields do not couple directly
to the dilaton. Thus, the lower dimensional ($D<10$) R-R fields $\chi$
and $B^{(2)}_{\mu\nu}$ are obtained from the ten-dimensional ones through
$\chi=\sqrt{V_{10-D}}\chi_{10}$ and $B^{(2)}=\sqrt{V_{10-D}}B_{10}^{(2)}$.

By combining the dilaton and the axion of the ten-dimensional type-IIB 
superstring into 
a single complex field $\lambda=\chi_{10}+ie^{-\phi_{10}}$, it is 
possible to check that the 
bosonic effective action is invariant under the $SL(2,{\bf R})$ 
transformation
$\lambda\rightarrow (a\lambda+b)/(c\lambda+d)$, $G_{\mu\nu}\rightarrow
|c\lambda+d|G_{\mu\nu}$ .\ Writing this transformation in 
terms of four-dimensional fields ($D=4$) we find that 
 the four-dimensional fields acquire an ``anomalous weight'' 
due to the fact
that $V_6$, as measured in the string frame, does transform under 
$SL(2,{\bf R})$ ( see the 3rd ref. of \cite{inhom}: 

\bea
\chi^{'}_{4}&=&{bd+ac\, e^{-2\phi_{4}}\over 
[(c\chi_{4}+d)^{2}+c^{2}e^{-2\phi_{4}}]^{1\over 4}}+
{ac\chi^{2}_{4}+(ad+bc)\chi_{4}\over [(c\chi_{4}+d)^{2}+
c^{2}e^{-2\phi_{4}}]^{1\over 4}
}, \nn\\ 
e^{-\phi'_{4}}&=&{e^{-\phi_{4}}\over [(c\chi_{4}+d)^{2}+
c^{2}e^{-2\phi_{4}}]^{1\over 4}}, \nn\\
H^{(1)'}&=&d\,H^{(1)}-c\,H^{(2)}, \nn\\
H^{(2)'}&=&[(c\chi_{4}+d)^{2}+c^{2}e^{-2\phi_{4}}]^{3\over 4}
(-b \,H^{(1)}+a\,H^{(2)}), \nn\\
G_{\mu\nu}^{'}&=&[(c\chi_{4}+d)^{2}+c^{2}e^{-2\phi_{4}}]^{1\over 2} 
G_{\mu\nu}\nn
\eea

\section{T-Duality}

The T-duality symmetry is probably one of the most interesting consequences of the 
string theory.  We often hear people say  that the transformation $R \longrightarrow 1/R$, where $R$ is the string
compactification scale, leaves the energy spectrum invariant. 
The existence of a compact direction is essential for this symmetry. I will not go here into great
details as to how this comes, just will make several remarks as applied to cosmology.

Suppose we probe by strings a 10 dimensional geometry, using two different
string theories. The topology of the 
10 dimensional space is 
$R^9  \times S^1$. The T-duality \cite{pol} identifies these two  string theories and, if $\alpha^{'}=\frac{1}{2\pi T}$, where
$T$ is the tension in both the theories, then the compactification radii satisfy $R_{1}R_{2}=
\alpha^{'}$. Note that if one of the radii shrinks to $0$ the other diverges. 

Compactification on a circle implies quantization of momenta which now comes in quantas of $1/R$.
These momenta appear as masses for states which are massless in higher dimensions.
In addition there exists yet another excitation type, the windings. If $m$ is the number the
string winds around the circle, the energy of these excitations is $m\times2\pi RT=
mR/\alpha^{'}$, therefore if the compactification radii are interchanged at the same time
as the momenta are interchanged with windings, the energy spectrum of the theory remains invariant.
Now, this is what happens on the level of the string theory. To relate this to the background spacetime where the strings
propagate, one must take into account that the 
strings propagate on a 10 dimensional target space, and in turn, the 
target space must satisfy that the so called $\beta$ functions are to vanish
( something roughly similar to Einstein equations, see for example \cite{horstief}).

If the target space happens to have a compact Killing direction then the 
actions we where talking about above are invariant under the following ``duality" transformation:
taking adapted coordinates in which 
$x^{0}$ denotes the coordinate along the Killing vector chosen to dualize, 
we find new values for $(G_{\mu\nu},\phi,B_{\mu\nu}^{(1)})$
\vskip 0.3in

$\begin{array}{*{20}c}
   {\tilde g\mathop {}\nolimits_{00}  = g\mathop {}\nolimits_{00} ^{ - 1} ,} & {\tilde g\mathop {}\nolimits_{0i}  = }  \\
\end{array}g\mathop {}\nolimits_{00} ^{ - 1} B\mathop {}\nolimits_{01} $
\vskip0.2in

$\tilde g\mathop {}\nolimits_{ij}  = g\mathop {}\nolimits_{ij}  - g\mathop {}\nolimits_{00} ^{ - 1} (g\mathop {}\nolimits_{i0} g\mathop {}\nolimits_{0j}  - B\mathop {}\nolimits_{i0} B\mathop {}\nolimits_{0j} )$

\vskip.2in

$\tilde B\mathop {}\nolimits_{ij}  = B\mathop {}\nolimits_{ij}  - g\mathop {}\nolimits_{00} ^{ - 1} (g\mathop {}\nolimits_{i0} B\mathop {}\nolimits_{0j}  - B\mathop {}\nolimits_{i0} g\mathop {}\nolimits_{0j} )$
\vskip.2in

$\begin{array}{*{20}c}
   {\tilde B\mathop {}\nolimits_{0i}  = g\mathop {}\nolimits_{00} ^{ - 1} g\mathop {}\nolimits_{0i} } & {\tilde \phi  = \phi  - 1/2\log g\mathop {}\nolimits_{00} }  \\
\end{array}$

\vskip.2in

The PBB cosmological models are based on the so-called scale factor duality which uses the 
T-duality symmetry of the string theory to invert the scale of expansion as in 
$\tilde g\mathop {}\nolimits_{00}  = g\mathop {}\nolimits_{00} ^{ - 1}$. In the case of FRW model, one may invert the 
overall scale factor due to isotropy of the model. What happens, then,  is that one may have 
a prolonged inflationary period in the inverted-scale model, while the standard model is decelerating.

{\it Excercise1}

Start with the solution of spatially flat FRW model driven by a massless scalar field.
Write it both in Einstein and in the String frame.
Use the T-duality transformations to obtain the model with an inverted scale factor.
What is the metric in the Einstein frame? What happens with
the scalar field? Find the kinematical quantities of the model ( acceleration, curvature scale etc).
What happens if you dualise this model just wrt a single Killing direction, say $\partial_{x}$? Are
the T-duality transformations invariant under diffeomorphism?

{\it Excercise2} \cite{fvm}

Start with a 4 dimensional  model 
\vskip 0.1in

$ds^2=-\left({2\eta\over t}-k\right)^{-\alpha}dt^2 + 
\left({2\eta\over t}-k\right)^{\beta} dx^2 + \left({2\eta\over t}-k\right)^{1-\alpha}
t^2[d\theta^{2}+f_{k}(\theta)^2 d\varphi^{2}]$
\vskip.2in
coupled to a dilaton field
\vskip.2in
$\phi={1\over 4}(\beta-\alpha)\log\left({2\eta\over t}-k\right).$

\vskip.2in
This is an exact solution of dilaton gravity provided the constants $\alpha$ and 
$\beta$ satisfy 
$$
\alpha^2+\beta^2=2.
$$ 
Here $k$ is the spatial curvature ($k=-1,0,1$) and 
$$
f_{k}(\theta)=\left\{\matrix{\sin{\theta} & k=1 \cr \theta & k=0 \cr
\sinh{\theta} & k=-1 } \right. .
$$

Identify all the Killing vectors. Show that only dualising wrt $\xi_{4}=\partial_{x}$ keeps
all isometries intact. Dualising wrt $\xi_{3}=\partial_{\varphi}$, for example, the homogeneity is lost,
yet the string theory remains invariant. This signals that the homogeneity, and isotropy
are not fundamental concepts of string cosmology. Identify the Schwarzschild solution among the above ( it has a constant scalar field) and speculate about the possible fate
of the black hole applying the duality transformation. 

\vskip.5in

There are some subtle differences  between 
scale factor duality and T-duality.
The inversion of one or more scale factors in the metric ( the scale factor duality) leaves 
the low-energy equations of
motion invariant. T-duality, is rather  stronger a tool, it is a transformation of the metric, not
necessarily only of the scale factors, leading to a new solution of the low-energy 
field equations which corresponds to an equivalent string theory. 
In the  metric of Exercise(2) we have four Killing vectors, and
 the degree of homogeneity may be reduced by 
just performing a T-duality transformation along $\phi$ direction. Consequently,
 for this model described by the line
element  with a cyclic $x$ coordinate, the only T-duality 
transformation leaving the spatial symmetry group intact is the one performed along the 
compactified isometric direction $x$.  In this case  it is formally identical to the scale 
factor duality of the metric. For scale factor duality, however, 
there is no need to impose the compactness of the $x$-coordinate since the ``dual'' model does 
not necessarily have to be equivalent to the original one as in complete string theories.

We have just only started to play the game and already on this level see that there are some interesting
aspects to string cosmology, for if for example the symmetries of the string theory are used
as a guide, we may conclude that two quite inequivalent backgrounds may host equivalent 
string theories and thus from the point of view of string propagation be indistinguishable.
Strings propagating on two backgrounds related by the T-duality transformation would not notice
the difference and wouldn't care of such aspects as homogeneity or isotropy.

Now, we have seen that the inversion of the scale factor is permitted by the string symmetries.
But what happens at $t=0$? there are different approaches to this problem, known in the 
literature as an `` exit problem". In the next section I will try to summarize our work on 
this problem.

\section{Regularisation of Singularities}

We all believe that the emergence of s-t singularities in
 GR suggests the breakdown of the theory at  natural scales of  the
 Theory.  
At these scales one expects the quantum corrections to 
take a leading role and save the situation.  While String/M theories are the
 best candidates to quantize gravity, near the singularity one can not use the
 perturbative approach, but rather a full-fledged M theory. Unfortunately we lack such a theory.
Different approaches to regularize singularity and to 
find a way out to solve the exit problem are discussed in the papers quoted in  \cite{sing}.

One of the ways of dealing with the problem, might be more qualitative an 
approach, trying to use the symmetries and dualities  of the M ( c.f. \cite{M}) theory to map singular
 backgrounds to nonsingular ones.
In plain words, one may formulate dualities in terms of physically equivalent vacua of the theory. 
Imagine one has two backgrounds A and B,
where every physical quantity of A may be written in terms of quantities of B. Now, if A is singular but B is not, we can say that the strings don't care, and the  singularity of A is  just irrelevant. This is to say that we were using the wrong degrees of freedom, or
 wrong ``tools" to describe the low energy limit of the theory.

	Another  hint that different degrees of freedom might be relevant 
comes already from the following simple \cite{gh} example. Imagine  there is a
 hidden extra dimension in the theory and take the scalar field spatially 
flat FRW cosmology in four dimensions.
It is  of course  singular at the ``Bang".
 Now, imagine this scalar field is just a ``trace" of an extra dimension. 
The 5 dim. lifted cosmology has a nonsingular  scalar curvature. This may hint that the 
4 dim. singularity is just an artifact of integrating the degrees of freedom 
associated with the extra dimension.
	The extra dimensions come naturally, or I would rather say are  {\it must}, 
if one considers S/M theory. So, let us try the two things:

1) Use the extra dimensions.

2) Use the M-theory symmetries, to get some info about singularities.

\vskip.1in

Our starting point would be a 4-D cosmology  with a bunch of massless 
minimally coupled scalar fields ( for more details see \cite{fvm2}.)  
Imagine we have designed an algorithm to construct such solutions. ( One can do it for quite a general class of models).
  We will concentrate on homogeneous scalar fields $\varphi_{i}(t)=q_{i}\varphi_{0}(t)$. Just think of these fields as 
all having the same functional dependence, but different amplitudes.

\vskip.1in

The field equations are, of course,  invariant under O(N) rotation between the fields.
 To be even more specific, let us look at the inhomogeneous generalization of  
spatially open FRW cosmology with N scalar fields.
\vskip0.1in
\bea
ds^2&=(\sinh{2t})^{\half(3\lambda-1)}(\cosh{4t}-
\cosh{4z})^{{3\over 4}(1-\lambda)}(-dt^2+dz^2)\cr 
&+\half\sinh{2t}\sinh{2z}\left(\tanh{z}\,dx^2+
{\rm cotanh\,}z\,dy^2\right)\nn
\eea

Here $\phi_{0}=\sqrt{3}/2\log\tan(t),$ and $\lambda$ is a sum of the squares of the amplitudes of the scalar
fields and is an $O(N)$ invariant.
The expression is written for generic $\lambda$ , 
 but in particular case $\lambda=1$  the model is just an open FRW solution  
( though in unusual co-ordinates) coupled to N scalar fields $\phi_{i}\sim p_{i}\log\tan(t)$                                                              .

Let us now ``lift" this solution to 4+N dimensions and see what happens.

The 4+N dimensional metric is:

\bea
ds^{2}_{4+N}&=2\,(\sinh{t})^{1-\sum_{i=1}^{N} p_{i}}
(\cosh{t})^{1+\sum_{i=1}^{N} p_{i}}
(-dt^2+dz^2+\sinh^2{z}\,dx^2+\cosh^{2}{z}\,dy^2)\cr
&+\sum_{i=1}^{N}
\tanh^{2p_{i}}{t}\,(dw^{i})^2\nn
\eea

The curvature invariant for the metric:

$$
R_{abcd}R^{abcd}\sim C(p_{i})t^{2(S-3)} +O[t^{2(S-2)}]
$$

where we have written $S=\sum_{i=1}^{N}p_{i}$ and $C(p_{i})$ is defined by 
$$
C(p_{i})={3\over 16}(S-1)^{2}(S^2-2S+5)+\sum_{i=1}^{N}
p_{i}^{2}[3+p_{i}^2+(S-3)(p_{i}+S)]+\sum_{i<j}^{N}p_{i}p_{j}
$$ 

The remarkable point about all this is that the Kretchman scalar $R_{abcd}R^{abcd}$ 
is regular whenever the condition $\lambda=1$ is satisfied. Thus, it may really be the case that
the singularity in 4 dimensions is just an artifact of integrating the extra degrees of freedom
encoded in higher dimensions.  

Let us now move forward and specify the number of extra dimensions to 7. In the 11 dimensional
geometry we can parametrize the moduli metric with the numbers $p_{i}$, which must satisfy the
following constraint:

\[
I = \sum\nolimits_1^7 {p\mathop {}\nolimits_i } ^2  + \frac{2}{3}\sum\nolimits_{i \ne j} {p\mathop {}\nolimits_i p\mathop {}\nolimits_j } 
\]

Now, we want to think of these solutions as representing vacua of the M theory. Therefore
we can look now at the $U$ duality group of M compactified on $(S^{1})^{7}$. It happens that
this group is generated by the so-called $\frac{2}{5}$ transformation which permutes the
$$
(p_{1},\ldots,p_{7})\rightarrow \left(p_{1}-{2s\over 3},p_{2}-{2s\over 3},
p_{3}-{2s\over 3}, p_{3}+{s\over 3},\ldots,p_{7}+{s\over 3}\right)
$$
with $s=p_{1}+p_{2}+p_{3}$.

These transformations map vacum solutions into vacuum solutions and one can check that 
singular solutions may be 
transformed by this transformation into no-singular ones.
The U-duality  transformations are conjectured to be an exact symmetry of the 
M-theory, and the fact that the two backgrounds, one singular, and one regular are connected via a 
U-duality transformation may indicate, as in the case of homogeneity and isotropy, that
certain kinds of singularities do not matter.
The results presented here is not a rigorous study, of course, nevertheless one can get 
an idea of what might happen near the singularity if the ideas o string cosmology apply.

Finally, I would like to present you with a scenario of a PBB universe where one starts with 
a small perturbation which collapses, inflates, gets rid of all its irregularities 
before approaching the intermediate singularity and after exiting approaches the standard model.

\section{A Universe from ``Almost Nothing"}

It is more or less accepted that to solve the flatness/horizon problem in cosmology
one must invoke the idea of accelerated expansion. There are two possibilities which provide
us with accelerated behaviour of the scale factor of the universe:

1) The standard inflation

2) The PBB scenario.

In the PBB picture one starts with the weakly coupled regime of the string theory and
its dynamics is controlled by the low energy effective action. This is in contrast with the standard inflation 
where one starts with physics on Plank scale and there is no way to extract a ``decent" Lagrangean
description for the inflaton field from a fundamental theory.  Moreover, the
mere existence of the dilaton coming from the string theory is quite ``unhealthy" for standard
picture of inflation.

To address the problem of initial conditions in PBB scenario, Bounanno, Damour and Veneziano (BDV) \cite{buon}
have formulated the basic postulate of ``asymptotic past triviality" identifying the 
initial state of the universe  with the generic perturbative solution of the tree-level low energy effective 
action. In this picture the initial state consists of a bath of gravitational and dilatonic
waves, some of which could have collapsed leading/modulo solution to exit problem to an FRW
universe. This picture generalizes and makes more concrete previous studies of inhomogeneous
versions of PBB cosmology \cite{jrd}, \cite{gr}.

However, within this generic approach one may perfectly well formulate and identify the problem, 
yet there are difficulties in resolving it. What we proposed \cite{fkvm1}, \cite{fkvm2}  is to probe the model with reduced difficulties, 
starting with strictly plane waves. Recently this work was generalized to higher dimensions and extra fields
in \cite{bv}.
 In this picture one solves the collapse/inflation problem analytically
relating the Kasner exponents near the singularity with the initial data for the dilaton and gravity
waves. Once we do so, the path is free to ask some phenomenological questions as to the 
conditions for successful inflation, or entropy generation in these models. 

We will be imposing the two BDV conditions in the distant past:

1) The string theory must be weakly coupled

$\exp(\varphi/2\ll 1)$

2) The curvature is small in string units.

To simplify the things we start with the graviton-dilaton system. 

$$
S=\int d^4 x\sqrt{-g}e^{-\phi }\left(R+g^{\alpha \beta }\partial _\alpha \phi
\partial _\beta \phi \right)
$$

Furthermore, we will assume throughout that the extra six spatial dimensions
are compactified in some internal appropriate manifold  considered 
to be non-dynamical.

The gravity wave being linearly polarized is specified by a single function $\psi(r,s)$ and
the scalar wave by $\phi(r,s)$, where $r$ and $s$ are null coordinates $r=t-z$ and $s=t+z$.

Both waves satisfy the same linear Euler Darboux equation  in the interaction region (c.f. \cite{szek})

\begin{eqnarray}
\psi_{,rs}+\frac{1}{2(r+s)}(\psi_{,r}+\psi_{,s})&=&0\nn\\
\phi_{,rs}+\frac{1}{2(r+s)}(\phi_{,r}+\phi_{,s})&=&0\nn
\end{eqnarray}

These two equations, together with the initial data on the null boundaries
of the interaction region $\{\psi(r,1),\psi(1,s)\}$
and $\{\phi(r,1),\phi(1,s)\}$
pose a well defined initial value problem. Both 
$\psi(r,s)$ and $\phi(r,s)$ are $C^{1}$ (and
piecewise $C^{2}$) functions.

Now, you may analytically integrate the solutions into the interaction region, and 
work out the Kasner exponents in terms of the incoming initial data.

The line element near the singularity is 

\begin{eqnarray}
ds^2=\xi^{a(z)}(-d\xi^2+dz^2)+\xi^{1+\epsilon(z)}dx^2
+\xi^{1-\epsilon(z)}dy^2
\nn
\end{eqnarray}
where $a(z)\equiv\frac{1}{2}[\epsilon^{2}(z)+ \varphi^{2}(z)-1]$

with
 
\begin{eqnarray}
\epsilon(z)&=&\frac{1}{\pi\sqrt{1+z}}
\int_{z}^{1}ds \left[(1+s)^{\frac{1}{2}}\psi(1,s)
\right]_{,s}
\left(\frac{s+1}{s-z}\right)^{\frac{1}{2}}\nonumber\\
&+&
\frac{1}{\pi\sqrt{1-z}}
\int_{-z}^{1}dr \left[(1+r)^{\frac{1}{2}}\psi(r,1)
\right]_{,r}
\left(\frac{r+1}{r+z}\right)^{\frac{1}{2}},
\nn
\end{eqnarray}
and a similar expression  for the dilaton
\begin{eqnarray}
\varphi(z)&=&\frac{1}{\pi\sqrt{1+z}}
\int_{z}^{1}ds \left[(1+s)^{\frac{1}{2}}\phi(1,s)
\right]_{,s}
\left(\frac{s+1}{s-z}\right)^{\frac{1}{2}}\nonumber\\
&+&
\frac{1}{\pi\sqrt{1-z}}
\int_{-z}^{1}dr \left[(1+r)^{\frac{1}{2}}\phi(r,1)
\right]_{,r}
\left(\frac{r+1}{r+z}\right)^{\frac{1}{2}}.\nn
\end{eqnarray}

The $\epsilon$ and $\varphi$ are the source functions expressed in terms of the
gravitational and scalar field respectively. So, for example, we may ask which are the models undergoing
the PBB inflation among the all possible solutions parametrized by $\epsilon$ and $\varphi$, and the answer
is `` the piece " of cake in the plane $\epsilon$ and $\varphi$, meaning that the initial data
leading to PBB is dense.

If the incoming data is weak ($\epsilon$ close to $0$ and $\varphi$
around $-1$)  the PBB inflation takes place . The fact that the inflationary data is dense one may say that
PBB is stable a feature of this collision.

How weak should the waves be to solve horizon/flatness problem?

One usually defines the $Z$ factor which is a ratio of co-moving Hubble radius at the
time when the Dilaton Driven Inflation starts to the Hubble radius at the time the picture breaks down either 
due to the strong coupling or to the strong curvature. One may work that the strong curvature regime,
and therefore the loop corrections are reached before the corrections in $\alpha^{'}$ become
important. Expressing the conditions of successful inflation  in terms of the focusing lengths
of the waves one finds:

$$
L_{1} L_{2} \stackrel{>}{\sim} e^{{60(\beta+2) \over 1-\alpha_{\rm min}}}\ell_{st}^{2}.
$$

$$
\beta\equiv \frac{2\varphi(z)}{a(z)+2},
$$
Here $L_{1}$ and $L_{2}$ are the focal lengths of the waves. 
Therefore the waves must be extremely weak to solve the flatness and horizon problems.
But this is nice, since this is what one expects anyway, if one has in mind a picture
of these waves as small perturbations on an otherwise flat background.

\subsection{Entropy Generation in a Wave-Collision-Induced PBB}

One of the interesting questions in this scenario is the way the entropy is 
generated in the collision. First, let me argue that the entropy in the plane wave region
is $0$ (both the matter and gravitational). Imagine you want to identify the states of $0$
gravitational entropy. First to mind comes Flat Space Time.  Then? Penrose suggested
FRW spacetime due to $0$  Weyl tensor. But if we take the string theories as a guide
the FRW  doesn't fit that well. Plane waves on the contrary represent the most simple exact \cite{horstief}
string backgrounds (1): {\it all the curvature invariants vanish}. So (2) if one would try to describe
the gravitational entropy with a homogeneous function vanishing at the origin ( not just
the Weyl tensor) one would choose the plane waves as $0$ entropy.

The 3rd point is due to triviality of plane wave s-t with respect to vacuum
polarization \cite{quant}: {\it no particle creation}! One usually relates the quantum particle 
production (gravitons) to gravitational entropy-no such production takes place in the vicinity of 
plane wave, stressing again the trivial entropy content.

What happens then, when two such waves collide? The  interaction region of the two incoming waves
can be divided into three regions. Just after the collision takes place at
$t_{i}$ the matter content of the universe can be satisfactorily described in terms of a 
superposition of the two incoming  non-interacting null fluids. In this regime the evolution is  approximately adiabatic and almost no
entropy is produced. This almost linear regime comes to an end
 as soon as the gravitational nonlinearities
 take over and the collision enters an intermediate phase where the dynamics of the
universe is dominated by both velocities and spatial gradients. Now the evolution 
is no longer adiabatic and matter entropy is generated. In this regime the matter 
content of the universe cannot be
described in terms of a perfect fluid equation of state \cite{taub} and some effective macroscopic 
description of the fluid, such as an anisotropic fluid or some other phenomenological 
stress-energy tensor, should be invoked . 

The production of entropy
will stop at the moment the velocities begin to dominate over spatial gradients and the 
evolution becomes  adiabatic again, the matter now being described by a perfect
fluid with stiff equation of state $p=\rho$. The transition to an adiabatic phase
will happen either before or at $t_{K}$ when the universe enters  the Kasner phase. 
{}From that moment on no further entropy is produced up to the end of DDI.
The relative duration of these three regimes crucially depends
on the strength of the incoming waves and on the initial data.
It is straightforward to show that for both 
gravitational and scalar waves the ratio of spatial gradients versus
time derivatives dies off as $1/\sqrt{L_1L_2}$  and therefore it takes the 
universe a time of order $\sqrt{L_1L_2}$ before the Kasner-like regime is reached, 
 fixing the  duration of the  nonadiabatic phase.

As discussed above, all the 
entropy is generated in the intermediate phase between the adiabatic  and 
Kasner regimes. On the other hand, in the Kasner regime the total entropy generated
in this intermediate region is carried adiabatically by the perfect stiff fluid 
represented by the dilaton field $\phi(t)$. If we consider a generic bariotropic 
equation of state, $p=(\gamma -1)\rho $, the first law of thermodynamics implies 
that the energy density $\rho$ and entropy density $s\equiv S/V$ evolve with the
temperature $T$ as 
\begin{eqnarray}
\rho &=&\sigma \,T^{\frac{\gamma }{\gamma -1}}  \nn \\
s &=&\gamma \,\sigma \,T^{\frac{1}{\gamma -1}},  \nn
\end{eqnarray}
where the temperature is
given by the usual relation $T^{-1}=(\partial s/\partial \rho )_{V}$.
Here $\sigma $ is a dimensionful constant which in the particular 
case of radiation ($\gamma=4/3$) is just the 
Stefan-Boltzmann constant. Using these equations   we easily
find that 
for a stiff perfect fluid ($\gamma =2$) the entropy density $s$ can be expressed in terms of
the energy density as 
\begin{eqnarray}
s=2\sqrt{\sigma\rho}. 
\nn
\end{eqnarray}
It is important to stress  that $\sigma$ is a parameter that
gives the entropy content of, and thus the number of degrees of freedom that
we associate with, the effective perfect fluid. In principle it
could be computed provided a microscopical description of the fluid is available. 
However in the case at hand, the stiff perfect fluid is just an effective description of
the classical dilaton field in the Kasner regime. Since the evolution in the DDI phase is
adiabatic, $\sigma$ can be seen as a phenomenological
parameter that measures the amount of entropy generated during the intermediate 
region where the dilaton field should be described by an imperfect effective
fluid. This effective description of the dilaton condensate, and the
entropy generated,  would then depend on a number of phenomenological parameters.

We can now give an expression for the total entropy inside a Hubble volume 
at the beginning of DDI in terms of the initial data. The energy density carried by
the dilaton field in the Kasner epoch is given by 
$$
\rho(t)={\dot{\phi}^2\over 4\ell_{Pl}^2}={\beta^2\over 4\ell_{Pl}^2 t^2}.
$$

Thus, the
total entropy inside a Hubble volume at the beginning of the Kasner regime 
can be written in terms of the Kasner exponents and the source function for the dilaton as
$$
S_{H}(t_{K})={t_{K}^{2}\over \ell_{Pl}^2}\sqrt{\sigma\ell_{Pl}^2}
{|\beta|\over \alpha_{1}\alpha_{2}\alpha_{3}}
$$
Now we can write $t_{K}=\eta \sqrt{L_{1}L_{2}}$ with 
$0<\eta\stackrel{<}{\sim} 1$, so finally
we arrive at
$$
S_{H}(t_{K})=\left[\eta^2\sqrt{\sigma\ell_{Pl}^2}{|\beta|\over \alpha_{1}\alpha_{2}\alpha_{3}}
\right]{L_{1}L_{2}\over \ell_{Pl}^2} 
\equiv \kappa {L_{1}L_{2}\over \ell_{Pl}^2}.
$$

It is interesting to notice that this scaling for the entropy in the
Hubble volume at 
the beginning of DDI can be also retrieved by considering a particular example when,
as a result of the wave collision,
a space-time locally isometric to that of a Schwarzschild black hole is produced
in the interaction region. In that case
the focal lengths of the incoming waves are related with the mass of the black hole
by \cite{dv} $M=\sqrt{L_{1}L_{2}}/\ell_{Pl}^{2}$. Now we can write
the Bekenstein-Hawking entropy of the black hole $S=4\pi\ell_{Pl}^{2}M^{2}$ in terms of
the focal lengths of the incoming waves as $S=4\pi L_{1}L_{2}/\ell_{Pl}^{2}$. 
This analogy is further supported by the fact that the temperature of the created 
quantum particles in both the black hole and the 
colliding wave space-time scales like \cite{temp} $T\sim 1/M$ and $T\sim 1/\sqrt{L_{1}L_{2}}$
respectively, as well as by the similarities between 
the thermodynamics of black holes and stiff fluids \cite{zurek}.
It is important to notice that 
this scaling of the temperature of the created particles    
with the focal lengths implies that, whenever enough inflation occurs,   
the contribution of these particles to the total entropy is negligible.

One may have thought, in principle,  that it is possible to avoid the
entropy production before the DDI by just taking a solution in the
interaction region for which the evolution is globally adiabatic. The
simplest possibility for such a solution would be a Bianchi I metric for
which the Kasner regime extends all the way back to the null boundaries.
This, however, should be immediately discarded due to the constraints 
posed by the boundary conditions in the colliding wave problem.
Or, put in terms of the null data, it is not possible to choose the
initial data on the null boundaries in such a way that the metric is
globally of Bianchi I type in the whole interaction region and at the same
time ${\cal C}^1$ and piecewise ${\cal C}^2$ across the boundary.

\section{Acknowledgements}
I am grateful to Nora Breton of the organizing committee for hospitality at Huatulco,
and to Kerstin Kunze and to Miguel Vazquez-Mozo for many hours of fruitful collaboration and discussions.


\vspace{.3in}
\begin{thebibliography}{99}
\bibitem{misner} C.W. Misner, Phys.Rev.Lett.{\bf 22},1071(1969)
\bibitem{guth} A. Guth, Phys.Rev.{\bf D23},347(1981)
\bibitem{penrose} R. Penrose, {\it Singularities and time-asymmetry}, in ``General Relativity,
an Einstein Centenary Survey''. Eds. S.W. Hawking and W. Israel, Cambridge University Press,
Cambridge, UK, 1979.
\bibitem{gabr} G. Veneziano, Phys. Lett. B {\bf 265}, 287 (1991);
M. Gasperini and G. Veneziano, Astropart. Phys.
{\bf 1}, 317 (1993); G. Veneziano, ``Inflating, 
warming up, and probing the pre-bangian universe'',
G. Veneziano, {\it String cosmology: The pre-big bang scenario}, in: ``The Primordial
Universe", proceedings to the 1999 Les Houches Summer School, eds. P. Binetruy, R. Schaeffer, 
J. Silk and F. David. Springer-Verlag 2001, hep-th/0002094;
E.J. Copeland, A. Lahiri and D. Wands, Phys. Rev.  {\bf D50},
4868 (1994); J.E. Lidsey, D. Wands and E.J. Copeland,
``Superstring cosmology''  Phys. Rep. {\bf337},  343 (2000)
 hep-th/9909061
for an updated collection of papers on string cosmology, see
{\rm http://www.to.infn.it/\~{ }gasperin}.
\bibitem{inhom}
G. Veneziano, Phys. Lett. {\bf B406},
297 (1997);  A. Buonanno, K.A. Meissner,  C. Ungarelli
and G. Veneziano , Phys. Rev. {\bf D57} (1998) 2543;
 A. Feinstein ,R.  Lazkoz  and   M. A. Vazquez-Mozo,
 Phys. Rev. {\bf D56} (1997) 5166; 
K. Saygili ,  Int. J. Mod. Phys. {\bf A14} (1999) 225; 
J. D.  Barrow  and  K. E. Kunze,  Phys. Rev. {\bf D56} (1997) 741;
ibid. {\bf D57} (1998) 2255; 
  D. Clancy,  A. Feinstein, J.E. Lidsey and  R. Tavakol, Phys. Lett.
{\bf B451} (1999) 303; 
Kunze K. E., Class. Quantum Grav. {\bf 16}, 286 (1999) gr-qc/9906073.
\bibitem{pol} Polchinski J.,
``String Theory", Cambridge University Press, 1998.
\bibitem{fvm} 
A. Feinstein and M.A. V\'azquez-Mozo, Phys. Lett. {\bf B441}, 40 (1998).

\bibitem{sing} R. Brustein  and  G. Veneziano , Phys. Lett. {\bf B329}  
(1994) 429; N. Kaloper, R. Madden  and K.A. Olive , Nucl. Phys. {\bf B452} (1995) 677,
Phys. Lett. {\bf B371}  34,(1996); R. Easther ,K.  Maeda and  D. Wands,  
Phys. Rev.{\bf D53} (1996) 4247.  M. Giovannini, Phys. Rev. {\bf D57} (1998) 7223;
R. Brustein  and  R. Madden,  Phys. Lett. {\bf B410,}, 110  (1997),
 Phys. Rev. {\bf D57} (1998) 712, JHEP 9907  006 (1999);
K.A. Meissner  and  G. Veneziano ,
 Mod. Phys. Lett. {\bf A6},  3397 (1991); M.  Gasperini, J. Maharana  and  G.  Veneziano, Phys.  
Lett. { \bf B296} 51 (1992),  M. Gasperini, J. Maharana and G. Veneziano, Nucl. Phys.  
{\bf B472} 349 (1996) ; 
 M. Gasperini  and Veneziano, G. Gen. Rel. Grav. {\bf 28} (1996) 1301; 
 A.A. Kehagias  and A. Lukas,  Nucl. Phys. {\bf B477} (1996) 549; 
A. Buonanno et al., Class. Quant. Grav. {\bf 14}  L97 (1997); M. Gasperini , M.  Maggiore and G. Veneziano ,
 Nucl. Phys. {\bf B494} (1997) 315; E.  Kiritsis and C. Kounnas , gr-qc/9509017, ``String Gravity and  
Cosmology: Some New Ideas", in Proc. of the Four Seas Conference, Trieste,  
1995.  M.  Maggiore and A. Riotto, Nucl. Phys. {\bf B548} 427  (1999) ;
A. Ghosh, R. Madden and G. Veneziano, Nucl.Phys. {\bf B570}, 207, hep-th/9908024. I.  Antoniadis,  J. Rizos and  K. Tamvakis, Nucl. Phys.  
{\bf B415} (1994) 497; R.  Easther and  K. Maeda, Phys. Rev. {\bf D54},  7252 (1996);
 R.  Brustein and  R. Madden, Phys. Rev. {\bf D57}, 712  (1998) ;  
  R. Brandenberger, R.  Easther and  J. Maia, JHEP {\bf 9808}, 007  (1998);
 S.  Foffa ,  M. Maggiore and  R. Sturani,  Nucl. Phys. {\bf B552}, 395  (1999);  
  C. Cartier, E. J. Copeland and  R. Madden, JHEP {\bf 0001}, 035  (2000), 
 hep-th/9910169.
 \bibitem{M} A. Lukas , B.A. Ovrut  and D. Waldram , Phys. Lett.  
{\bf B393},65 (1997)  Nucl. Phys. {\bf B495},365 (1997);
 F.  Larsen and F. Wilczek , Phys. Rev. {\bf D55}, 4591 (1997) ;
 N. Kaloper, Phys. Rev. {\bf D55} , 3394 (1997) ;
  H. Lu, S. Mukherji and Pope, Phys. Rev. {\bf D55},7926  (1997) ;
 R.   Poppe and  S. Schwager, Phys. Lett.
{\bf B393} , 51 (1997) ;
  A. Lukas and  B. A. Ovrut, Phys. Lett. {\bf B437}, 291  (1998) ;
  N. Kaloper,  I.  Kogan and  K. A. Olive, Phys. Rev. {\bf D57}, 7340  (1998);
T. Banks ,W.  Fischler and  L. Motl, JHEP {\bf 9901} 019 (1999).

\bibitem{gh} G.W. Gibbons and P.K. Townsend, Nucl. Phys. {\bf B282} 610 (1987);
G.W. Gibbons, G.T. Horowitz and P.K. Townsend, Class. Quant. Grav. {\bf 12} 297 (1995)

\bibitem{fvm2} A. Feinstein  and M.A. Vazquez-Mozo, Nucl. Phys. {\bf B568}, 405 (2000), hep-th/9906006.

\bibitem{buon} A. Buonanno, T. Damour and G. Veneziano,
Nucl. Phys.  {\bf B543}, 275 (1999).

\bibitem{jrd}
D. Clancy, J.E. Lidsey and R. Tavakol, Phys. Rev. {\bf D58}, 044017 (1998);
Phys. Rev. {\bf D59}, 063511 (1999).
\bibitem{gas} M. Gasperini, Phys.Rev. {\bf D61}, 087301  (2000),
gr-qc/9902060 .
\bibitem{gr} 
D. Clancy, A. Feinstein, J.E. Lidsey and R. Tavakol, Phys. Rev. {\bf D60}, 043503 (1999);
Phys. Lett. {\bf B451}, 303 (1999); K.E. Kunze, Class. Quantum Grav. {\bf 16}, 3795 (1999). 

\bibitem{fkvm1}  A. Feinstein, K.E. Kunze and M.A. V\'{a}zquez-Mozo, 
 Class. Quantum Grav.{\bf 17}, 3599 (2000), hep-th/0002070.

\bibitem{fkvm2}  A. Feinstein, K.E. Kunze and M.A. V\'{a}zquez-Mozo, Phys. Lett. {\bf B491} 190 (2000) 
\bibitem{szek}  P. Szekeres, J. Math. Phys. {\bf 13}, 286 (1972);
A. Feinstein and J. Ib\'{a}\~nez, Phys. Rev. 
{\bf D39}, 470 (1989);
U. Yurtsever, Phys. Rev. {\bf D38}, 1706 (1988); P. Schwarz, Nucl. Phys. {\bf B373},
529 (1992).

\bibitem{bv} V. Bozza and G.Veneziano, JHEP {\bf 0010},  035 (2000)

\bibitem{horstief}
D. Amati, C. Klim\v{c}ik, Phys. Lett. {\bf B219}, 443 (1989);
R. G\"uven, Phys. Lett {\bf B191}, 275 (1987);
G.T. Horowitz and A.R. Steif, Phys. Rev. Lett. {\bf 64}, 260 (1990).
\bibitem{quant}
S. Deser, J. Phys. {\bf A8}, 1972 (1975);
G. Gibbons, Commun. Math. Phys. {\bf 45}, 191 (1975);
J. Garriga and E. Verdaguer, Phys. Rev. {\bf D43}, 391 (1991);
O. Levin and A. Peres, Phys. Rev. {\bf D50}, 7421 (1991);
A. Feinstein, M.A. P\'erez-Sebastian, Class. Quantum Grav. {\bf 12}, 2723 (1995).
\bibitem{taub}  R. Tabensky and A. H. Taub, Commun. Math. Phys. {\bf 29}, 61 (1973)
\bibitem{dv} M. Dorca and E. Verdaguer, Nucl. Phys. B {\bf 
403}, 770 (1993).
\bibitem{temp} U. Yurtsever, Phys. Rev. {\bf D40}, 360 (1989);
A. Feinstein and M.A. P\'erez-Sebastian,
Class. Quantum Grav.  {\bf 12}, 2723 (1995).
\bibitem{zurek} W.H. Zurek and D.N. Page, Phys. Rev. {\bf D29}, 628
(1984),
A. Feinstein, {\it The stiff state, entropy bounds and the black hole
 thermodynamics},
BGU-preprint-1985 (unpublished)
\end{thebibliography}
\end{document}